\documentclass[10pt,journal,cspaper,compsoc]{IEEEtran}
\ifCLASSOPTIONcompsoc
\else
  \usepackage{cite}
\fi
\ifCLASSINFOpdf
  \usepackage[pdftex]{graphicx}
  \DeclareGraphicsExtensions{.pdf,.jpeg,.png}
\else
\fi
\usepackage[cmex10]{amsmath}
\usepackage{amssymb}
\usepackage{array}
\usepackage{tabularx}
\usepackage{multirow}
\usepackage{slashbox}
\usepackage{booktabs}
\usepackage{subfigure}
\usepackage{algorithm}    
\usepackage{algorithmic}

\numberwithin{equation}{section}

\begin{document}

\title{Attributes Coupling based Item Enhanced Matrix Factorization Technique for Recommender Systems}

\author{Yonghong~Yu, Can Wang,
        Yang~Gao,~\IEEEmembership{Member,~IEEE},

\IEEEcompsocitemizethanks{\IEEEcompsocthanksitem Y. Yu and Y. Gao (Corresponding Author) are with the State Key Laboratory for Novel Software Technology, Nanjing University, Nanjing,
        210093, China. \protect (e-mail: yuyh.nju@gmail.com and
gaoy@nju.edu.cn).\protect\\

\IEEEcompsocthanksitem Can Wang is with the CSIRO Computational Informatics, Australia. \protect (e-mail: canwang613@gmail.com)

}

\thanks{Manuscript received xxx xx, 201x; revised xxx xx, 201x. This work
was supported in part by the National Science Foundation of China under Grants
61035003,  61175042,  and  60721002, the National 973 Program of
China under Grant  2009CB320702, the 973 Program of Jiangsu, China under Grant
BK2011005, and the Program for New Century Excellent Talents in University under
Grant NCET-10-0476.}}

\markboth{IEEE TRANSACTIONS ON KNOWLEDGE AND DATA ENGINEERING, VOL. xx,
NO. xx, xxx 201x}%
{YU \MakeLowercase{\textit{et al.}}: Attributes Coupling based Item Enhanced Matrix Factorization Technique for Recommender Systems}

\IEEEcompsoctitleabstractindextext{%
\begin{abstract}
Recommender system has attracted lots of attentions since it helps users alleviate the information overload problem. Matrix factorization technique is one of the most widely employed collaborative filtering techniques in the research of recommender systems due to its effectiveness and efficiency in dealing with very large user-item rating matrices. Recently, based on the intuition that additional information provides useful insights for matrix factorization techniques, several recommendation algorithms have utilized additional information to improve the performance of matrix factorization methods. However, the majority focus on dealing with the cold start user problem and ignore the cold start item problem. In addition, there are few suitable similarity measures for these content enhanced matrix factorization approaches to compute the similarity between categorical items. In this paper, we propose attributes coupling based item enhanced matrix factorization method by incorporating item attribute information into matrix factorization technique as well as adapting the coupled object similarity to capture the relationship between items. Item attribute information is formed as an item relationship regularization term to regularize the process of matrix factorization. Specifically, the similarity between items is measured by the \emph{Coupled Object Similarity} considering coupling between items. Experimental results on two real data sets show that our proposed method outperforms state-of-the-art recommendation algorithms and can effectively cope with the cold start item problem when more item attribute information is available.
\end{abstract}

}

\maketitle

\section{Introduction}
\IEEEPARstart{R}{ecommender} systems \cite{adomavicius2005toward} are intelligent software tools that provide web users with the decision-making support information, such as what movie to watch, what book to read and what product to buy. At present, recommender systems have become indispensable since it overcomes the information overload problem, by providing web users with the personalized information, products or services to satisfy their tastes and preferences. In this paper, we unify these information, products and services and call them  `items'. In order to keep customer loyalty and prompt sale revenues, more and more e-commerce sites deploy recommender systems to meet users' information demands. Some typical web applications equipped with recommender systems include product recommendation in Amazon\footnote{http://www.amazon.com}, radio recommendation in Last.fm\footnote{http://www.Last.fm}, movie recommendation in Netflix\footnote{http://www.netflix.com} and friend recommendation in LinkedIn\footnote{https://www.linkedin.com} etc.

Collaborative filtering (CF) \cite{breese1998empirical,resnick1994grouplens} is one of the most widely used techniques for building recommender systems and has achieved great successes in E-commerce for its domain independency (i.e. collaborative filtering only requires the past activities history of users to make recommendations, and does not depend on the types of Items).
However, collaborative filtering suffers from the following limitations \cite{adomavicius2005toward,su2009survey}.

\begin{enumerate}
  \item Data Sparsity. A modern E-commerce recommender system may include millions of users and millions of items. Even a very active user, however, exhibits a relatively small proportion of items available in E-commerce systems. Meanwhile, even the very popular items are rated by only a tiny part of users existing in E-commerce systems. Facing the sparsity of available user activity records, it is difficult for collaborative filtering based recommender systems to discover similar users or similar items according to their rating behaviors. As a result, the collaborative filtering based recommender systems are unable to generate personalized  recommendations for users. This problem, in general, referred to as the data sparsity problem, is the major issue that leads to negative effects on the recommendation quality of the collaborative filtering based recommender systems.
  \item Cold Start Problem.  Cold start problem can be categorized into cold start user problem and cold start item problem. Cold start users refer to the users who have just joined the e-commerce system and have expressed very few ratings. Hence, the collaborative filtering based recommender systems are incapable to provide accurate recommendations for cold start users, due to the lacking of sufficient rating information to find cold start users' neighbors or learn their latent preferences. Similarly, cold start items refer to new items or items that only have received a small number of ratings from users. Hence, cold start items cannot be
   accurately recommended until they have been rated by a sufficient number of users.
  \item Scalability. In order to make recommendations for users, recommender systems equipped with traditional collaborative filtering algorithms need to compute the pairwise similarities among users or among items, whose time complexity of computing similarities grows exponentially with the number of users and the number of items. As the rapidly growing amount of users and items available in E-commerce systems, traditional collaborative filtering algorithms suffer seriously from scalability problems.
\end{enumerate}

Many work has been proposed to overcome different types of issues mentioned above in the research of recommender systems. For instance, in order to deal with the data sparsity issue,  Sarwar et al. \cite{sarwar1998using} and Yongli Ren \cite{ren2013lazy} adopted imputation techniques to filling the missing ratings and make the user-item rating matrix dense. However, data imputation is still in its infancy and several issues involved data imputation still remain unexplored, such as how to select the most important missing data to fill in. On the other hand, several clustering techniques based recommendation algorithms have been proposed to cope with the scalability issue. Rashid et al. proposed CLUSTKNN \cite{al2006clustknn}, which uses a variant of basic k-means algorithm to partition users into clusters, and then leverages a CF algorithm to produce recommendations. Xue et al. proposed CBSMOOTH [12], which uses the clusters as the computed groups and smoothes the unrated data for individual users. Although clustering techniques based recommendation algorithms can improve the scalability of recommender systems, they often provide less personalized recommendations and often lead to poor accuracy. To overcome cold start problems, earlier work combined traditional collaborative filtering with user demographics or product descriptions to alleviate the cold start problem \cite{melville2002content,yu2004probabilistic,ziegler2004taxonomy}, more recently research concentrates on extending the matrix factorization method \cite{zhen2009tagicofi,ma2008sorec,jamali2010matrix}, to which our work belongs to.

In the last years, matrix factorization \cite{koren2009matrix} methods have drawn lots of attentions due to their good scalability and predictive accuracy. In addition, matrix factorization technique offers a flexible framework to incorporate additional sources of information to improve the recommendation quality. Moreover, Koren \cite{koren2009matrix} and Adomavicius \cite{adomavicius2005toward}, who both are famous research scientists in the research of recommender system,  argued that additional information, such as social network information, user demographics and item descriptions, may provide useful
information for matrix factorization technique to improve the recommendation performance. Following by the hints and with more rich additional sources of information become available, several recommendation approaches are introduced to extend the matrix factorization techniques by utilizing additional information recently. For example, in \cite{zhen2009tagicofi}, Zhen Yi et al. proposed TagiCoFi to seamlessly integrate tagging history into the matrix factorization framework. Hao Ma et al. \cite{ma2008sorec} and Jamali et al. \cite{jamali2010matrix} present social recommendation algorithms based on matrix factorization by employing both users' social network information and rating records. Their experimental results demonstrate that those additional information can be leveraged to improve the recommendation quality.

Various additional information has been exploited to improve the quality of recommendation under the matrix factorization framework. However, the majority focus on dealing with the cold start user problem by leveraging all kinds of additional information and ignore the cold start item problem, which our work try to tackle by leveraging item attribute information with matrix factorization framework.

Item attribute information is an important supplement to the user interaction records and has been exploited to improve the performance of recommendation algorithms. For instance, Kim et al. \cite{kim2004probabilistic} incorporated item attributes into a item-based probabilistic model to solve the cold start item problem. Hence, we can inherit the advantages of matrix factorization approach as well as cope with the cold start problem by combining matrix factorization approach and item attribute information.

To the best of our knowledge, there exists only one recommendation algorithm \cite{nguyen2013content} which attempts to combine matrix factorization approach and item attribute information to improve the recommendation quality. Specifically, in this method, the similarity between different items is measured by the simple matching similarity ( SMS ) \cite{gan2007data}, which is too
rough to capture the closeness of two items.



In this paper, we propose attributes coupling based item enhanced matrix factorization method by incorporating item attribute information to
 overcome the cold start item problem, and consequently improve the quality of recommendation. Specifically, item attribute information is exploited to regularize the matrix factorization by adding item relationship regularization term to the objective function of matrix factorization. The item relationship regularization term makes two item-specific latent feature vectors as similar as possible if the two items have similar attribute contents. Furthermore, in order to deeply capture the relationship between items, \emph{Coupled Object Similarity} (COS) \cite{wang2011coupled,cao2012coupled} is adapted to measure the interactions or couplings between items. The effectiveness of COS in capturing genuine relationships between items described by categorical attributes has been validated in \cite{wang2011coupled,yu2013coupled}. Experimental results on two real-life data sets show that our proposed method outperforms the state-of-art recommendation methods, and can effectively cope with the cold start item problem when more item attribute information is available.

 The key contributions of our work are summarized as follows:
 \begin{itemize}
   \item we propose attributes coupling based item enhanced matrix factorization method. By combining item attribute information and matrix factorization framework, we can cope with the cold start item problem existed in matrix factorization, and at the same time, inherit the advantages of matrix factorization approach.
   \item we capture the relationships among items based on COS, which has been evaluated to outperform other similarity measures (e.g., SMS \cite{gan2007data},ADD \cite{ahmad2007k}) for categorical data. By this means, we overcome the similarity measure problem in matrix factorization framework.
   \item we perform extensive experiments to evaluate our proposed method on two real data sets in terms of the recommendation quality and the effectiveness of tackling the cold start item problem.
 \end{itemize}


The rest of this paper is organized as follows. Section \ref{sec:relatework} briefly reviews related work in recommender systems. Section \ref{sec:pre} introduces the preliminary knowledge used in this paper. Section \ref{sec:itembased} describes the details of our proposed item recommendation algorithm by combining matrix factorization framework with item attribute information, whose relationships are measured by the coupled object similarity metric. Experiments are evaluated in Section \ref{sec:experiments}. Finally, we conclude this paper and present some directions for future work in Section \ref{sec:conclusion}.

\section{Related Work}
\label{sec:relatework}
Collaborative filtering (CF) \cite{adomavicius2005toward,breese1998empirical,resnick1994grouplens,su2009survey} approaches have achieved a great success in the research of recommender systems since CF methods are domain independent and only require the past activities history of users, i.e. user-item rating matrix, to make recommendations. According to different means of utilizing the user-item rating matrix, collaborative filtering approaches can be divided into two main categories \cite{breese1998empirical}: memory-based algorithms and model-based algorithms.

Memory-based filtering algorithms, also known as neighbor-based methods, use the entire user-item rating matrix to generate recommendations. Memory-based methods firstly employ various similarity measures to find user neighborhood or item neighborhood for the active user or target item, respectively. Once the neighborhoods are formed, memory-based filtering algorithms usually take a weighted sum of ratings given by their neighbors (active user' neighbors or target item' neighbors) as a prediction for target item. Typical memory-based algorithms include user-based methods \cite{resnick1994grouplens,breese1998empirical} and item-based methods \cite{sarwar2001item,linden2003amazon}. User-based approaches predict the ratings based on the opinions of active user's neighbors, which have similar preferences with active user. On the other hand, item-based approaches provide predictions based on the ratings given by active user for items similar to target items in terms of rating patterns.

In contrast with memory-based filtering approaches, which utilize entire user-item matrix to provide recommendations for active users, model-based filtering approaches first make use of statistical and machine learning techniques to learn a predictive model from training data. The predictive model can characterize the rating behaviors of active users. Then model-based filtering approaches use the trained model to make predictions, rather than directly utilize the entire user-item matrix to compute predictions. Typical examples of model-based filtering approaches include Bayes networks \cite{breese1998empirical}, clustering model \cite{ungar1998clustering,xue2005scalable,yu2013coupled}, latent semantic analysis \cite{hofmann2004latent,hofmann2003collaborative}, restricted boltzmann machines \cite{salakhutdinov2007restricted} and association rules \cite{sarwar2000analysis,lin2002efficient}. Breese et al. \cite{breese1998empirical} presented a collaborative filtering algorithm based on Bayesian networks learned from training data. Hofmann et al. \cite{hofmann2004latent} introduces latent class variables to discover user communities and prototypical interest profiles. Ungar et al. \cite{ungar1998clustering} grouped similar users in the same class and make predictions according to active user's neighbors belonged to the same class with active user. Sarwar et al. \cite{sarwar2000analysis} applied association rule discovery algorithms to seek association between co-purchased items and then provided recommendations based on the strength of the association between items.

Generally, memory-based algorithms tend to easy to implement and produce reasonable highly prediction quality. However, memory-based algorithms suffer from serious scalability problem. As the volume of of user and item sets increasingly grow, their worse online performance  make it not appropriate for modern E-commerce sites. Model-based algorithms tend to be faster than memory-based algorithms in terms of response time. The disadvantages of model-based algorithms are that many theoretical models are complex and  are not fit well with real data. In addition, it takes a long time to build or update models for model-based algorithms.

Since the great success of Netflix Prize competition, matrix factorization \cite{koren2009matrix} based recommendation algorithms have gained great popularity due to their effectiveness and efficiency in dealing with very large user-item rating matrix. Based on the assumption that only a few factors contribute to a user's preference and item's characteristics, matrix factorization approaches simultaneously embed both user and item feature vectors into a low dimension latent factor space, where the correlation between user's preference and item's characteristics can be computed directly, and then utilize their low dimension representations to make further recommendations. Examples of matrix factorization based recommendation algorithms include Singular Value Decomposition (SVD) \cite{sarwar2000application}, Nonnegative Matrix Factorization (NMF) \cite{seung2001algorithms,cai2011graph,zhang2006learning}, Maximum-Margin Matrix Factorization (MMMF) \cite{srebro2004maximum,rennie2005fast}, Probabilistic Matrix Factorization (PMF) \cite{mnih2007probabilistic}, nonparametric matrix factorization (NPCA) \cite{yu2009fast}.

The above mentioned matrix factorization methods for recommender systems only utilize user-item rating information to learn latent user feature vector and item feature vector, but ignore additional information, for instance, social networks, tagging information and item attribute information etc.. Although the proceeding matrix factorization methods can effectively and efficiently deal with large user-item rating information, they may fall into cold start problem since the sparsity of user-item rating information.

Recently, based on the intuition that additional information may be useful for improving the performance of recommender systems, especially for overcoming the cold start user problem, several matrix factorization algorithms have been proposed. For example, Zhen Yi et al. \cite{zhen2009tagicofi} proposed TagiCoFi to seamlessly integrate tagging history into the matrix factorization framework. Le Wu \cite{wu2012leveraging} proposed a two-stage recommendation framework, named as Neighborhood-aware Probabilistic Matrix Factorization (NHPMF), to improve recommendation accuracy. The NHPMF extended the probabilistic matrix factorization method by leveraging tagging data. Hao Ma et al. \cite{ma2008sorec} and Jamali et al. \cite{jamali2010matrix} proposed social recommendation algorithms based on matrix factorization by employing both users' social network information and rating information.
These extensions of matrix factorization methods leverage additional information, such as tagging data and social relations, to infer the similarity among users. Then the preprocessed similarity information are incorporated into some kind of basic matrix factorization methods to guarantee that the learned latent user feature vectors are close as possible to that of neighbors of users. These approaches are specially effective for tackling cold start user problem and force the latent feature vectors of new user with no or very few ratings to depend on the latent feature vector of their most similar neighbors whose latent feature vectors can be accurately learned from user-item matrix.

However, there are several problems with these methods. First, Tagging data, expressed as words, are labeled by user arbitrarily. Taking social relations as the similarity between users, which is too coarse-grained to distinguish the degree of similarity between different users since the similarity value take 1 only if two users have trust relationship, otherwise 0. Moreover, they only consider cold start user problem and ignore the cold start item problem.

In contrast, item attribute information, for example, \emph{director, actor, genre} for movie item, generated by domain experts, can more accurately represent the characteristics of item. Hence, item attributes information can be exploited to deal with cold start item problem and improve the quality of recommendation. However, few work focus on exploiting item attributes information to improve the quality of recommendation. To the best of knowledge, only Nguyen et al. \cite{nguyen2013content} proposed content-boosted matrix factorization method for recommender systems by utilizing item attribute content to improve recommendation quality. In the content-boosted matrix factorization method, the similarity between two items is measured according to the simple matching similarity, which is too rough to capture the genuine relationships among items.

\section{Preliminary Knowledge}
\label{sec:pre}
In this section, we introduce the preliminary knowledge related to our proposed attributes coupling based item enhanced matrix factorization algorithm.
We first introduce the notations used in this paper in Section \ref{sec:notations}. Then, in Section \ref{sec:mf}, we briefly describe the matrix factorization based recommendation algorithm. Finally, we present the Coupled Object Similarity (\emph{COS}) \cite{wang2011coupled,yu2013coupled}, which is used to measure the relationships among items based on item attributes information.

\subsection{Notations}
\label{sec:notations}
In a typical scenario, a recommender system consists of a set of $N$ users $U=\{ u_1,u_2,...,u_N\}$, and a set of $M$ items $I=\{i_1,i_2,...,i_M \}$. Generally, user preferences on items are usually converted into a user-item rating matrix $R$, with $N$ rows and $M$ columns. Each entry $r_{ui}$ of $R$ represents the rating given by user $u$ on item $i$. In principle, $r_{ui}$ can be any real number, but usually ratings are integers and fall into [0,5], in which $0$ indicates that the user has not yet rated that item. A higher rating corresponds to better satisfactory. The set of items rated by the user $u$ is denoted as $I_u$($I_u\subseteq I$).

In practical, the user-item rating matrix $R$ is generally very sparse with many unknown entries since a typical user may have only rated a tiny percentage of items. For example, in MovieLen100K data set and Netflix data set,
93\% and 99\% of the possible ratings are missing, respectively. Consequently, the sparse nature of user-item rating matrix leads to poor recommendation quality.

Moreover, each item $i \in I$ is represented as an attribute vector $a_i=\{a_{i_1},a_{i_2},...,a_{i_D}\}$, where $D$ is the number of attributes. These attribute vectors are extracted from content information of items, and they are categorical in nature. For example, if the item set $I$ represents a collection of movies, then the attributes, i.e., \emph{director, actor, genre}, are extracted to express a movie item. In addition, those attributes have categorical values, such as ``\emph{Drama}'',``\emph{War}'' and ``\emph{Comedy}'' etc. for the attribute \emph{genre}. All item attribute vectors form item-attribute information matrix $A$, and each entry $a_{ij}$ of $A$ represents the value of attribute $a_j$ for item $i$.

In essence, the objective of recommender systems is to predict the rating on the specified item $i$ for an active user $u$, denoted by $\widehat{r}_{ui}$, by leveraging all available sources of information by all kinds of machine learning techniques.

\subsection{Matrix Factorization for Recommender Systems}
\label{sec:mf}
Matrix factorization technique is widely employed in the research of recommender systems. The goal of matrix factorization technique is to learn the latent preferences of users and the latent characteristics of items from all known ratings, then predict the unknown ratings through the inner products of user latent feature vectors and item latent feature vectors. Formally, matrix factorization based methods decompose the user-item rating matrix $R$ into two low rank latent feature matrices $P \in \mathbb{R}^{K\times N}$ and $Q \in \mathbb{R}^{K\times M}$, where $K\ll min(N,M)$, and then use the product of $P$ and $Q$ to approximate the rating matrix $R$. As a result


\begin{equation} \label{eq:1}
R \approx \widehat{R} = P^{T}Q=\begin{bmatrix} p_{1}^{T}\\ p_{2}^{T} \\...\\p_{N}^{T} \end{bmatrix}\begin{bmatrix} q_{1} \quad q_{2} \quad ... \quad q_{M} \end{bmatrix}
\end{equation}
The column vectors $p_{u}$ and $q_{i}$ represent the $K$-dimensional user-specific latent feature vector and item-specific latent feature vector, respectively. Once recommender systems gain the low rank latent feature matrices, we can use the inner product of $p_{u}$ and $q_{i}$ to estimate the rating given by the active user $u$ for target item $i$. Formally,
\begin{equation} \label{eq:2}
\widehat{r}_{ui} = p_{u}^{T}q_{i}
\end{equation}

In order to learn the latent feature vectors of users and items, we solve the approximate problem described above in a traditional way by utilizing the Singular Value Decomposition (SVD) \cite{paterek2007improving}, which minimizes the following objective function,

\begin{equation} \label{eq:3}
\frac{1}{2}  \parallel  R-P^{T}Q \parallel ^{2} _{F}
\end{equation}
where $\parallel . \parallel ^{2} _{F}$ is the Frobenius norm \cite{golub2012matrix}. Although SVD is a powerful technique for identifying latent semantic factors in information retrieval, it is not well-defined when the user-item rating matrix is highly sparse. Hence, it is common to directly factorize the observed ratings only and turn objective function (\ref{eq:3}) into


\begin{equation} \label{eq:4}
\min_{p^*,q^*} \frac{1}{2} \sum_{(u,i)\in T}(R _{u i }-p^{T}_u q_i ) ^{2}
\end{equation}
where $T$ indicates the set of the $(u,i)$ pairs for known ratings. To avoid over-fitting, two regularization terms on the sizes of $P$ and $Q$ are added into Equation (\ref{eq:4}). As a result, Equation (\ref{eq:4}) is changed to
\begin{equation} \label{eq:5}
\ell=\min_{p^*,q^*} \frac{1}{2} \sum_{(u,i)\in T}(R _{u i }-p^{T}_u q_i ) ^{2} +\frac{\lambda_1}{2}\parallel P \parallel ^{2} _{F} + \frac{\lambda_2}{2}\parallel Q \parallel ^{2} _{F}
\end{equation}
where $\lambda_1,\lambda_2$ represent the regularization parameters and control the impacts on the learnt latent feature vectors.

Due to both $P$ and $Q$ being unknown, the optimization problem in Equation (\ref{eq:5}) is biconvex. Usually, an efficient and easy-to-implementation algorithm called the stochastic gradient descent algorithm (SGD) \cite{nemirovski2009robust} is applied to seek a local minimum solution of the objective function given by Equation (\ref{eq:5}). The SGD algorithm keeps on iterating on the training set until the objective function shown in Equation (\ref{eq:5}) converges to or arrivals at the upper bound of the number of iterations.

To learn the user latent feature matrix $P$, we fix $Q$. Then the derivative of $\ell$ with respect to $p_u$ is as follows,
\begin{equation} \label{eq:6}
\frac{\partial \ell}{\partial p_u} = -(R _{u i }-p^{T}_u q_i)q_i + \lambda_1 p_u
\end{equation}

Similarly, we learn the item latent feature matrix $Q$ by firstly keeping $P$ fixed. Then the derivative of $\ell$ with respect to $q_i$ is displayed below,
\begin{equation} \label{eq:7}
\frac{\partial \ell}{\partial q_i} = -(R _{u i }-p^{T}_u q_i)p_u + \lambda_2 q_i
\end{equation}
Accordingly, the stochastic gradient descent algorithm uses the following updating rules to learn the latent feature vectors $p_u$, $q_i$:
\begin{equation} \label{eq:8}
p_u \longleftarrow p_u + \eta( (R _{u i }-p^{T}_u q_i)q_i - \lambda_1 p_u )
\end{equation}
\begin{equation} \label{eq:9}
q_i \longleftarrow q_i + \eta( (R _{u i }-p^{T}_u q_i)p_u - \lambda_2 q_i )
\end{equation}
where $\eta$ is the learning rate.

The matrix factorization algorithm described above is the so-called Regularized Singular Value Decomposition (RSVD) \cite{paterek2007improving}, which is widely employed due to its good scalability and high recommendation quality. From the perspective of Bayesian, RSVD is equivalent to Probabilistic Matrix Factorization \cite{mnih2007probabilistic}, which has been demonstrated to be one of the state-of-the-art collaborative filtering methods.

In this paper, we take the RSVD method as a baseline approach and enhance it by incorporating item attributes information to improve the recommendation quality and make recommendations more interpretable.
\subsection{Item Relationship Measure \emph{COS}}
\label{sec:cos}
In recommender systems, items are usually described by categorical attributes. For example, a movie item can be represented by a collection of categorical features (i.e. director, actor, genre and country). There are few suitable similarity measures to compute the similarity between items described by categorical attributes. For instance, in  \cite{nguyen2013content}, which is one of our main comparison algorithms, Nguyen et al. use simple matching similarity to measure the closeness between items $i$ and $i'$. Formally,
\begin{equation} \label{eq:17}
SMS(i,i^{'})=\frac{\sum_{j=1}^D\delta(a_{ij},a_{i^{'}j})}{D}
\end{equation}
where $D$ is the number of attributes. $\delta(a_{ij},a_{i^{'}j})$ is the simple match similarity between $a_{ij}$ and $a_{i^{'}j}$ and is defined as follows,
\begin{equation} \label{eq:smsdeta}
\delta(a_{ij},a_{i^{'}j})=
\begin{cases}
1& \text{if $a_{ij} = a_{i^{'}j} $}\\
0& \text{if $a_{ij} \neq a_{i^{'}j} $}
\end{cases}
\end{equation}

In essence, for categorical data, the SMS only uses $0$ and $1$ to distinguish similarities between distinct and identical categorical values. Hence, it is relatively rough and fails to capture the genuine relationship between categorical data. For example, by using the simple matching similarity measure, the similarity between two items described as [`A1',`B1',`C1'] and [`A1',`B2',`C2'] is 0.33, while this similarity based on Table 2 in \cite{wang2011coupled} by using the coupled object similarity is 0.75, which more accurately reflects the relationship between categorical data.

Therefore, we adopt the Coupled Object Similarity (\emph{COS})$\in$ [0,1] proposed in \cite{wang2011coupled} to measure the similarity between items based on the item-attribute information matrix $A$. The \emph{COS} considers both the intra-coupled similarity within an attribute and the inter-coupled similarity between different attributes, where the effectiveness of \emph{COS} in capturing genuine relationship between items described by categorical data has been validated in \cite{wang2011coupled,yu2013coupled}.

Formally, the Coupled Object Similarity (\emph{COS}) between categorical items $i$ and $i'$ is defined as follows.
\begin{equation}
COS(i,i')=\sum_{j=1}^n\delta^A_j(a_{ij},a_{i'j})
\label{eq:COS}
\end{equation}
where  $a_{ij}$ and $a_{i'j}$ are the values of attribute $a_j$ for $i$ and $i'$, respectively; and  $\delta^A_j$ is Coupled Attribute Value Similarity( \emph{CAVS}) between attribute values $a_{ij}$ and $a_{i'j}$.

The \emph{CAVS} $\delta^A_j$ consists of the \emph{Intra-coupled Attribute Value Similarity (IaAVS)} measure $\delta_{j}^{Ia}(a_{ij},a_{i'j})$ and the \emph{Inter-coupled Attribute Value Similarity (IeAVS)} measure  $\delta_{j}^{Ie}(a_{ij},a_{i'j})$ for attribute $a_j$. The \emph{IaAVS} takes value occurrence frequency
within an attribute into account and reflects the value similarity in terms of frequency distribution, while the \emph{IeAVS} considers the dependency aggregation among attributes and reflects the value similarity in terms of item value co-occurrence. By simultaneously considering both \emph{IaAVS} and \emph{IeAVS}, the definition of \emph{CAVS} between attribute values $a_{ij}$ and $a_{i'j}$ is as follows.
\begin{equation}
\delta^{A}_{j}(a_{ij},a_{i'j})=\delta_{j}^{Ia}(c)\cdot\delta_{j}^{Ie}(a_{ij},a_{i'j})
\label{eq:CAVS}
\end{equation}

In detail, based on the intuition that more similar occurrence frequencies of an attribute value pair indicate greater similarity and higher occurrence frequencies of an attribute value means more importance \cite{gan2007data}, the \emph{Intra-coupled Attribute Value Similarity (IaAVS)} measure $\delta_{j}^{Ia}(a_{ij},a_{i'j})$ is defined as follows:
\begin{equation}
\delta_{j}^{Ia}(a_{ij},a_{i'j}) =\frac{|g_j(a_{ij})|\cdot|g_j(a_{i'j})|}{|g_j(a_{ij})|+|g_j(a_{i'j})|+|g_j(a_{ij})|\cdot|g_j(a_{i'j})|}
\label{eq:IaAVS}
\end{equation}
where $g_j(a_{ij})$ and $g_j(a_{i'j})$ are the set information functions, which denote the set of items that their values of attribute $a_j$ are $a_{ij}$ and $a_{i'j}$, respectively.

On the other hand, the \emph{Inter-coupled Attribute Value Similarity  (IeAVS)} measure $\delta_{j}^{Ie}(a_{ij},a_{i'j})$ between attribute values $a_{ij}$ and $a_{i'j}$ can be computed by:
\begin{equation}
\delta_{j}^{Ie}(a_{ij},a_{i'j})=\sum_{k=1,k\neq j}^D \alpha_k\delta_{j|k}(a_{ij},a_{i'j})
\label{eq:IeAVS}
\end{equation}
where $\alpha_k$ is the weight of attribute $k$, $\alpha_k\in[0,1]$, all $\alpha_k$ sums up to $1$, and $\delta_{j|k}(a_{ij},a_{i'j})$  is defined as:
\begin{equation}
\delta_{j|k}(a_{ij},a_{i'j})=\sum_{w\in\bigcap}\min\{P_{k|j}(\{w\}|a_{ij}),P_{k|j}(\{w\}|a_{i'j})\}
\label{eq:IRSI}
\end{equation}
where $\bigcap$ denotes the intersection set of $\varphi_{j\rightarrow k}(a_{ij})$ and $\varphi_{j\rightarrow k}(a_{i'j})$, whose elements are values of attribute $a_k$ for items that their values of attribute $a_j$ are $a_{ij}$ and $a_{i'j}$, respectively. $P_{k|j}(\{w\}|x)$ is the information conditional probability of attribute value $w$ with respect to another attribute value $x$ and is defined as follows:
\begin{equation}
P_{k|j}(w|x)=\frac{|g_k(w)\bigcap g_j(x)|}{|g_j(x)|}
\end{equation}

Overall, by adopting the coupled object similarity to measure the similarity among categorical items, we can accurately capture the genuine relationship among items and better characterize the item latent feature vectors in the process of matrix factorization, hence produce more accurate recommendations compared to conventional approaches.

\section{Attributes Coupling Based Item Enhanced Matrix Factorization Method}
\label{sec:itembased}
In this section, we propose our attributes coupling based item enhanced matrix factorization method for recommender systems, in which item attribute information are utilized to regularize the matrix factorization procedure.

\subsection{Framework of Attributes Coupling Based Item Enhanced Matrix Factorization Method}
\label{sec:framework}

The key idea of our proposed recommendation algorithm is to utilize item attribute information to regularize the matrix factorization. The item attribute information is formed as an item relationship regularization term and makes an assumption that two item latent feature vectors $q_i$ and $q_{i'}$ are similar if the two items have similar characteristics in terms of item attribute information.


In order to make two item latent feature vectors $q_i$ and $q_{i'}$ as similar as possible if they are relatively close according to their item attribute contents, we add an item relationship regularization term based on item attribute information to constrain the baseline matrix factorization framework, i.e. RSVD. The item relationship regularization term is defined as:
\begin{equation} \label{eq:10}
 \frac{\beta}{2}\sum_{i=1}^M \sum_{i'=1}^M S_{i,i'}\parallel q_i - q_i'  \parallel ^{2} _{F}
\end{equation}
where $\beta$ is another regularization parameter to control the impact from the item attribute information, $S_{i,i'}$ is the similarity between two items based on their item attribute information. The similarity between  items $i$ and $i'$ forms the $(i,i')$ entry of similarity matrix $S$. In our proposed approach, this similarity is measured by using coupled object similarity \cite{wang2011coupled,cao2012coupled}, which has been described in Section \ref{sec:cos}. A small value of $S_{i,i'}$ means that the distance of two item latent feature vectors must be great, while a small value of distance indicates that $S_{i,i'}$ must be large. Hence, this term relationship regularization term makes two item latent feature vectors more ``close'' if they share some common characteristics based on their item attribute information.


Let Q be expressed as $[q_1,q_2,...,q_M]$ and $e_i$ indicate the element column vector, then $q_i=Q e_i$. We can rewrite item relationship regularization term as follows:


\begin{equation} \label{eq:11}
\begin{aligned}
 & \frac{\beta}{2}\sum_{i=1}^M \sum_{i'=1}^M S_{i,i'}\parallel q_i - q_i'  \parallel ^{2} _{F} \\   & =\frac{\beta}{2}\sum_{i=1}^M \sum_{i'=1}^M S_{i,i'}\parallel Q e_i - Q e_{i'}  \parallel ^{2} _{F}\\
 & = \frac{\beta}{2}\sum_{i=1}^M \sum_{i'=1}^M S_{i,i'} (Q(e_i-e_{i'}))^{T}Q(e_i-e_{i'}) \\
 & = \frac{\beta}{2}\sum_{i=1}^M \sum_{i'=1}^M S_{i,i'} tr((Q(e_i-e_{i'}))^{T}Q(e_i-e_{i'})) \\
 & = \frac{\beta}{2} tr( Q \sum_{i=1}^M \sum_{i'=1}^M \{ (e_i-e_{i'}) (e_i-e_{i'})^T S_{i,i'}\}  Q^T)    \\
 & = \frac{\beta}{2} tr( Q \mathcal{L} Q^T)    \\
\end{aligned}
\end{equation}
where $\mathcal{L}= D-S $ represents the Laplacian matrix and $D$ is a diagonal matrix with diagonal elements $D_{ii}= \sum_{i'=1}^M S_{i,i'}$.

By adding the item relationship regularization term into Equation \ref{eq:5}, our proposed attributes coupling based item enhanced matrix factorization method can be formulated as:


\begin{equation} \label{eq:12}
\begin{split}
\ell^* &=\min_{p^*,q^*} \frac{1}{2} \sum_{(u,i)\in T}(R _{u i }-p^{T}_u q_i ) ^{2} +\frac{\lambda_1}{2}\parallel P \parallel ^{2} _{F} \\
 & + \frac{\lambda_2}{2}\parallel Q \parallel ^{2} _{F} + \frac{\beta}{2}\sum_{i=1}^M \sum_{i'=1}^M S_{i,i'}\parallel q_i - q_i'  \parallel ^{2} _{F}
 \end{split}
\end{equation}
Replacing Equation \ref{eq:10} with Equation \ref{eq:11}, we can change the objective function \ref{eq:12} to

\begin{equation} \label{eq:13}
\begin{split}
\ell^* &=\min_{p^*,q^*} \frac{1}{2} \sum_{(u,i)\in T}(R _{u i }-p^{T}_u q_i ) ^{2} \\ & +\frac{\lambda_1}{2}tr(P^TP) + \frac{\lambda_2}{2}tr(Q^TQ) + \frac{\beta}{2}tr( Q \mathcal{L} Q^T)
\end{split}
\end{equation}
Similar to the RSVD approach, we seek a local minimum solution of the objective function derived from Equation \ref{eq:13} by applying the stochastic gradient descent algorithm. To learn the latent feature vectors, we use the following updating rules for $p_u$ and $q_i$:

\begin{equation} \label{eq:14}
p_u \longleftarrow p_u + \eta[ (R _{u i }-p^{T}_u q_i)q_i - \lambda_1 p_u ]
\end{equation}
\begin{equation} \label{eq:15}
\begin{split}
q_i  \longleftarrow q_i + \eta[ (R _{u i }-p^{T}_u q_i)p_u - \lambda_2 q_i \\
                  -\beta (q_i\sum_{i'=1}^M S_{i i'} - \sum_{i'=1}^M S_{i i'} q_{i'} )]
\end{split}
\end{equation}
From updating Equations \ref{eq:14} and \ref{eq:15}, it is easy to see that the gradient with respect to $p_u$ is identical to Equation \ref{eq:6}, while the gradient with respect to $q_i$ changes to

\begin{equation} \label{eq:16}
\frac{\partial \ell^*}{\partial q_i} = -(R _{u i }-p^{T}_u q_i)p_u + \lambda_2 q_i + \beta (q_i\sum_{i'=1}^M S_{i i'} - \sum_{i'=1}^M S_{i i'} q_{i'} )
\end{equation}
To summarize, our proposed attributes coupling based item enhanced matrix factorization approach for recommender system is described in Algorithm \ref{alg:1}.

\renewcommand{\algorithmicrequire}{\textbf{Input:}}
\renewcommand{\algorithmicensure}{\textbf{Output:}}

\begin{algorithm}[H]
\caption{Attributes Coupling Based Item Enhanced Matrix Factorization for Recommender System}
\label{alg:1}
\begin{algorithmic}[1]
\REQUIRE ~~\\
$\textbf{R}$ : the user-item rating matrix.\\
$\textbf{K}$ : the dimension of latent feature vector.\\
$\textbf{A}$ : the item-attribute information matrix.\\
$W$ : the number of iterations.\\
$\lambda_1$ : the regularization parameter for user regularization term.\\
$\lambda_2$ : the regularization parameter for item regularization term.\\
$\beta$ : the regularization parameter for item relationship regularization term.\\
$\eta$ : the learning rate.\\
\ENSURE ~~\\
$\textbf{P}$: the user latent feature matrix.\\
$\textbf{Q}$: the item latent feature matrix.\\
\STATE Compute the item similarity matrix $\textbf{S}$ by utilizing the \emph{COS} metric based on item-attribute information matrix $\textbf{A}$.
\STATE Initialize $\textbf{P}^0$,$\textbf{Q}^0$ with random decimals and $j=0$.

\WHILE{$j < W$ or [$\ell^{*(j)}- \ell^{*(j+1)} \leq \epsilon $]}
\FOR{ $u=1$ to $N$ and $i=1$ to $M$}
\IF{$R_{ui}\neq 0$}
\STATE Compute the gradient with respect to $p_u$ and $q_i$ by using Equations \ref{eq:6} and \ref{eq:16}, respectively.
\STATE Update $p_u$ and $q_i$ with Equations \ref{eq:14} and \ref{eq:15}.
\ENDIF
\ENDFOR
\STATE $j$++.
\ENDWHILE
\RETURN $\textbf{P}$ and $\textbf{Q}$.
\end{algorithmic}
\end{algorithm}

\subsection{Complexity Analysis}
\label{sec:complexity}
In our proposed recommendation algorithm, the main computation cost involves two parts: learning the latent feature vectors and computing the similarity among items with the coupled object measure.

The main computation cost of learning parameters is to evaluate the objective function $\ell^*$ and its gradients against latent user and item feature vectors. The computational complexity of evaluating objective function is $O(M\overline{r}K+M\overline{t}K)$, where $\overline{r}$ is the average number of ratings per item and $\overline{r}$ is the average number of most similar neighbors per item. Since the user-item rating matrix $R$ is extremely sparse, the value of $\overline{r}$ is relatively small. On the other hand, in our proposed matrix factorization model, we always choose items that are most similar to target item as the neighbors of target item, which indicates that $\overline{t}$ generally takes relatively small value. Hence, the computation of $\ell^*$ is fast and linear with respect to the number of items $M$ in the user-item rating matrix $R$. Assuming the average number of ratings per user is $\overline{x}$, the time complexities of evaluating $\frac{\partial \ell^*}{\partial p_u}$ and $\frac{\partial \ell^*}{\partial q_i} $ are $O(N\overline{x}K)$ and $O(M\overline{r}K+M\overline{t}^{2}K)$, respectively. Hence, the total time complexity of computing the gradients in each iteration is $O(M\overline{r}K+M\overline{t}^{2}K)$.

The overhead of computing the similarity among items is $O(D^{2}\rho^{3})$, where $\rho$ is the maximal number of attribute values for all the attributes in item-attribute information matrix $A$. Generally, the value of $D$ is small. For instance, the value of $D$ is 1 in MovieLens data sets\footnote{http://grouplens.org/datasets/movielens/}, in which only the attribute genre for movie items is available. In extreme case, the famous Netflix data set \footnote{http://www.netflixprize.com/} even has no attribute information about items. To the best of our knowledge, only HetRec2011 data set \footnote{http://ir.ii.uam.es/hetrec2011/datasets.html} includes relatively complete item attribute information, in which $D=4$. In contrast, the value of $\rho$ often takes a large value, which leads to an expensive computation overhead.  Since the process of computing similarity among items is offline, it does not add any additional cost to the process of learning latent feature vectors.

\section{Experiments and Evaluation}
\label{sec:experiments}
In this section, we conduct several experiments on real data sets to compare the performance of our proposed attributes coupling based item enhanced matrix factorization method, referred as to \emph{IEMF}, with other state-of-the-art methods. We address the following questions.
\begin{enumerate}
  \item How does our proposed \emph{IEMF} compare with other state-of-the-art collaborative filtering approaches, especially with matrix factorization technique based recommendation algorithms?
  \item How does the control parameter $\beta$ and $K$ impact the quality of recommendation?
  \item Can \emph{IEMF} effectively tackle the cold start item problem?
  \item How does the size of item's neighborhood affect the recommendation results?
\end{enumerate}

\subsection{Data Sets Description}
\label{sec:dataset}
Several data sets have been widely used to evaluate the performance of recommendation algorithms, such as Movielens, EachMovie\footnote{http://grouplens.org/datasets/eachmovie/}, Netflix, Epinions\footnote{http://www.epinions.com}, HetRec2011 etc.. However, only Movielens100K and HetRec2011 contain item attribute information. Hence, we choose these two data sets (i.e. MovieLens100K and HetRec2011) to evaluate our proposed method.

MovieLens100K contains 100,000 ratings from 943 users and 1,682 movies. Users with less than 20 ratings have been removed. The sparsity level of MovieLens100K is 1- $\frac{100000}{943*1682}$, which is equal to 93.69\%.
HetRec2011 is an extension of MovieLens10M, published by GroupLens research group. HetRec2011 contains 855,598 ratings given by 2,113 users on 10,197 movies. The sparsity level of HetRec2011 is 96.03\%, which is sparser than MovieLens100K. Moreover, Movielens100K lacks director, actor, and country etc. attributes and only contains genre information, while HetRec2011 data set includes relatively complete item attribute information and contains director, actor, country and genre attributes. In our experiments, we extract director, country and genre etc. attributes from HetRec2011 data set to represent the item attribute vectors and only extract genre attribute to describe the item attribute vectors in Movielens100K.

Note that, the original HetRec2011 data set is incomplete. For instance, some movie items do not have country attribute, others do not have director attribute, even the genres of several movie items are incorrectly labeled.  Moveover, the numbers of movie items are not consistent among movie item information that include country attribute and that include director attribute information as well as that contain genres information. So, we preprocess the HetRec2011 data set before our experiments by leveraging the IMDB\footnote{http://www.imdb.com/}.

General statistics about these two data sets are summarized in Table \ref{tab:dataset}.
\begin{table}[!hbp] \renewcommand{\arraystretch}{1.3}
\centering
\small{
\caption{\textbf{Statistics of Moivelens100K and HetRec2011}}
\begin{tabular} {c|c|c}
\toprule
\textbf{ Statistics }& \textbf{Moivelens100K} & \textbf{HetRec2011} \\ \hline
Num. of Ratings & 100,000 & 851,871   \\  \hline
Num. of Users, $N$ & 943 & 2,113   \\   \hline
Num. of Items, $M$ & 1,682 & 10,046  \\   \hline
Sparsity & 0.9369 & 0.9599  \\   \hline
Avg. Ratings per User & 106.04 & 403.157  \\   \hline
Avg. Ratings per Item & 59.45 & 84.80  \\
\bottomrule[1pt]
\end{tabular}
 \label{tab:dataset}}
 \vspace{-0.4cm}
\end{table}

\begin{figure}[!h]
    \begin{center}
        \includegraphics[scale=.64]{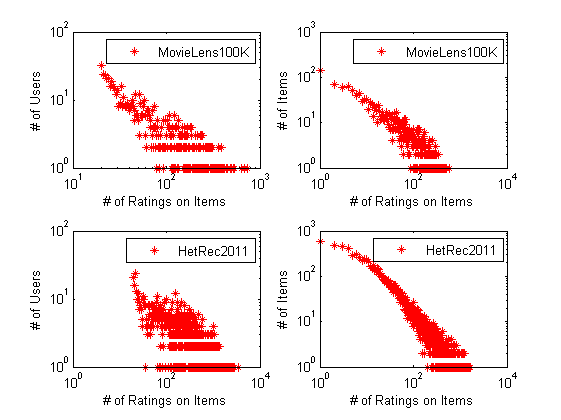}
    \end{center}
    \caption{Power-law distribution of two datasets.}
    \label{fig:powlaw}
\end{figure}

In Fig. \ref{fig:powlaw}, we also plot the power distributions of two data sets. From Figure \ref{fig:powlaw}, we can observe that the number of ratings per item shows more serious long tail effect that of per user for both data sets. In other words, the negative effect of cold start items is larger than that of cold start users on recommendation quality. This difference hints us that we should pay more attention to the cold start item problem than to the cold start user problem, which is the motivation of our proposed method.

\subsection{Evaluation Metrics}
\label{sec:metric}
We choose two popular metrics:  \emph{Mean Absolute Error (MAE)} and \emph{Root Mean Squared Error (RMSE)}, to measure the recommendation quality of our proposed method compared with other recommendation algorithms. Formally,
 \begin{equation}
MAE=\frac{\sum_{i=1}^T |p_i-q_i|}{T}
\label{eq:MAE}
\end{equation}
\begin{equation}
RMSE=\sqrt{\frac{\sum_{i=1}^T |p_i-q_i|^{2}}{T}}
\label{eq:RSME}
\end{equation}
where $p_i$ and $q_i$ are the real rating and the corresponding prediction, respectively, and $T$ denotes the total number of predictions generated for all active users.

From above equations, we can see that the lower the \emph{MAE} or \emph{RMSE}, the better the recommendation algorithm.

\subsection{Compared Approaches and Experimental Settings}
\label{sec:setting}
In order to evaluate the performance of our proposed method, we choose the following state-of-the-art approaches for comparison.
\begin{enumerate}
 \item \emph{\textbf{RSVD}}: RSVD is proposed by Arkadiusz Paterek \cite{paterek2007improving}. This method learns latent feature vectors by minimizing the sum-of-squared error between real ratings and estimations for available ratings in training set. It has been demonstrated to be one of the state-of-the-art collaborative filtering methods and only utilizes user-item rating matrix $R$ to generate recommendations.
  \item \emph{\textbf{NMF}}: This method is proposed by Lee et al. \cite{lee1999learning,seung2001algorithms}. Different from other matrix factorization techniques, it adds one more constraint on matrix factor model: both low rank latent feature matrices $P$ and $Q$ only have positive entries. This method also utilizes user-item rating matrix $R$ to produce recommendations.
  \item \emph{\textbf{PMF}}: This method is represented by Salakhutdinov et al. \cite{mnih2007probabilistic} and can be viewed as a probabilistic extension of the SVD model. \emph{PMF} represents the latent user and item feature vector by means of a probabilistic graphic model with Gaussian observation noise. Similar to \emph{RSVD} and \emph{NMF}, \emph{PMF} learns the latent user and item feature vector only based on rating information.
  \item \emph{\textbf{CBMF}}\cite{nguyen2013content}: This method is proposed by Nguyen et al. \cite{nguyen2013content}. To facilitate comparison, We refer this method as \emph{CBMF}. \emph{CBMF} incorporates content information directly into the matrix factorization approach to improve the quality of recommendation. More specially, the simple matching similarity metric is used to measure the relationship between two categorical items.
\end{enumerate}

In order to make a fair comparison, we set the common parameters to be identical parameter values in all methods.  For all involved recommendation algorithms, we set $\lambda_1=\lambda_2=0.1$. Meanwhile, the learning rate $\eta$ in all methods is set to be 0.005. Specially, the control parameters $\beta$ in\emph{ CBMF} and \emph{IEMF} are set to 0.1. Finally, we use $\epsilon = 0.0001$ and the number of iteration $W=200$ to control the loop conditions of matrix factorization procedures.

We conduct a five-fold cross validation over Moivelens100K and HetRec2011 data sets by randomly extracting different training and test sets at each time, which accounts for 80\% and 20\%, respectively.  Finally, we report the average results on test sets.

We use a PC with a Intel Xeon CPU@3.2GHz Processor, 8GB memory, Windows2003 Server operating system and J2SE 1.7, to conduct all our experiments.

\subsection{ Recommendation Quality Comparisons}
\label{sec:quality}

\begin{table*}[!h] \renewcommand{\arraystretch}{1.3}

\centering
\small{
\caption{Recommendation Quality Comparisons }
\begin{tabular}  {c||c|c|c|c|c|c|c}
\toprule[1pt]
 Dataset & Dimension $K$ & Metric & \emph{RSVD} & \emph{NMF} & \emph{PMF} & \emph{CBMF} & \emph{IEMF} \\ \toprule[1pt]
\multirow{4}{*}{Movielens100K} & \multirow{2}{*}{10} & \emph{MAE} & 0.7468& 0.7919 & 0.7519 & 0.7308 & \textbf{0.7282} \\
\cline{3-8}
 & & \emph{RMSE }& 0.9576 & 1.0027 & 0.9663 & 0.9213 & \textbf{0.9186} \\
\cline{2-8}
&\multirow{2}{*}{50}  & \emph{MAE} & 0.7437 & 0.7774 & 0.7674 & 0.7298 & \textbf{0.7277} \\
\cline{3-8}
 & & \emph{RMSE} & 0.9594 & 0.9840 & 0.9757 & 0.9198 & \textbf{0.9182}  \\
 \bottomrule[1pt]
 \multirow{4}{*}{HetRec2011} & \multirow{2}{*}{10} & \emph{MAE} & 0.6091  & 0.6287 & 0.6082 & 0.6026 & \textbf{0.5802} \\
\cline{3-8}
 & & \emph{RMSE }& 0.7910 & 0.8317 & 0.8000 & 0.7845 & \textbf{0.7667} \\
\cline{2-8}
&\multirow{2}{*}{50}  & \emph{MAE} & 0.6178 & 0.6355 & 0.6159 & 0.6097 & \textbf{0.5920} \\
\cline{3-8}
 & & \emph{RMSE} & 0.8234 & 0.8308 & 0.8219 & 0.7922 & \textbf{0.7816}  \\
 \bottomrule[1pt]
\end{tabular}
 \label{tab:PQresult} }
 \vspace{-0.4cm}
\end{table*}
Table \ref{tab:PQresult} reports the results of recommendation quality for the above selected recommendation algorithms, in which the number of dimensions $K$ of latent feature vectors are set to be 10 and 50.

From Table \ref{tab:PQresult}, we can observe that approaches \emph{CBMF} and \emph{IEMF} outperform other methods, which only utilize the user-item rating matrix to learn latent feature vectors. \emph{CBMF} improves the MAE of \emph{PMF} by  2.8\% and 1\% on MovieLens100K and HetRec2011 with $K=10$, respectively. With the same parameters settings, our proposed method \emph{IEMF} improves the MAE of \emph{PMF} by 3.2\% and 4.6\% on MovieLens100K and HetRec2011, respectively.

This observation confirms the assumption that using item content information can improve the recommendation quality. Moveover, for \emph{CBMF} and \emph{IEMF}, which both integrate item content information into matrix factorization to improve the recommendation quality, \emph{IEMF} generally achieves better result than \emph{CBMF} on both data sets. This observation demonstrates that our \emph{COS} measure is more accurate than \emph{SMS} in capturing the genuine relationship between two categorical items. Hence, \emph{COS} measure is more helpful to generate better recommendations.

It should be noted that on the HetRec2011 data set, MAE and RMSE values generated by the above selected methods when $K=50$ are generally higher than the corresponding values when $K=10$, which means that a high dimension of latent feature vectors may degrade the performance of recommendation algorithms based on matrix factorization technique. A possible reason is that continuously increasing $K$ may introduce noise into the matrix factorization model after $K$ arrivals at the optimal value to characterize the user and item features adequately.

In addition, all our selected methods perform better on HetRec2011 than on Moivelens100K. This is due to the fact that the average number of ratings per user in HetRec2011 is much larger, which is nearly 4 times the corresponding number of MovieLens100K. Moveover, the gain of our proposed method \emph{IEMF} over \emph{NMF} on HetRec2011 in term of MAE is greater than the gain on Moivelens100K. This phenomena indicates that our proposed method can work better when more attribute information are available since HetRec2011 includes country, director, actor and genre attributes, while MovieLens100K only contains genre attribute.

\subsection{Impact of Control Parameter $\beta$}
\label{sec:beta}
In our proposed method, the parameter $\beta$ plays an important role and controls the influence of item attribute information on learning the item latent feature vectors. A larger value of $\beta$ indicates that we put more weights on item attribute information to predict items' characteristics. In the extreme case, item attribute information would dominate the learning process and make item latent feature vectors close to its direct neighbors. A small value of $\beta$ makes our method degrade to baseline \emph{RSVD} method. Hence, very large values of $\beta$ or very small values of $\beta$ hurt the recommendation quality.
In this section, we perform a group of experiments to evaluate the impact of $\beta$ on the performance of our proposed method by changing the values of $\beta$ from 0.01 to 1. Another parameter $K$ is set as $K=10$.
\begin{figure}[!h]
    \begin{center}
        \includegraphics[scale=.5]{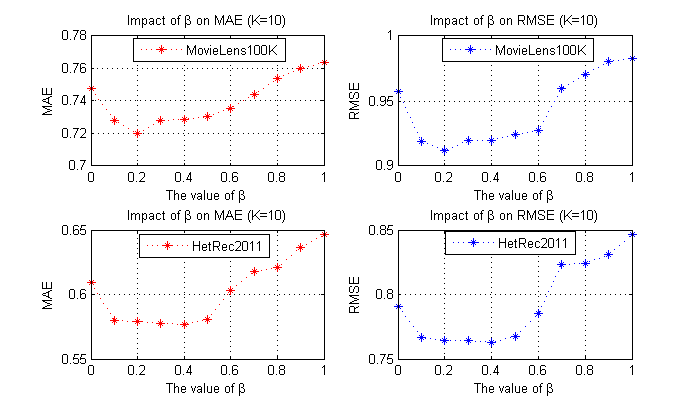}
    \end{center}
    \caption{Impact of Different $\beta$ on MAE and RMSE.}
    \label{fig:beta}
\end{figure}

Fig. \ref{fig:beta} reports the impacts of parameter $\beta$ on MAE and RMSE for both data sets. From Figure \ref{fig:beta}, we have the following observations: (1) the values of $\beta$ have a significant impact on the recommendation quality, which indicates that combining user-item rating information and item attribute information can greatly improve the recommendation quality, (2) the curves of MAE and RSME on two data sets show similar change trends.  As the $\beta$ increases, the values of MAE and RMSE firstly drop down, the recommendation quality improves, after the parameter $\beta$ reaches a certain threshold, the MAE and RMSE begin to increase as the parameter $\beta$ increases, which means that the performance degrades when $\beta$ is too large. This observations indicate that only using user-item rating matrix by abandoning item attribute information or excessively rely on item attribute information cannot generate reliable recommendations.

Moveover, our recommendation approach achieves the best performance: MAE=0.7197 when $\beta$ is around 0.2 on MovieLens100K, while we get MAE=0.5765 at $\beta=0.4$ on HetRec2011. This phenomenon demonstrates that our recommendation approach with HetRec2011 depends more on item attribute information than that with MovieLens100K, which confirms that more available item attribute information is helpful for alleviating the cold start item problem since HetRec2011 contains a larger portion of cold start items than that of MovieLens100K.

\subsection{Impact of Dimension of Latent Feature $K$}
\label{sec:K}
\begin{figure}[!h]
    \begin{center}
        \includegraphics[scale=.5]{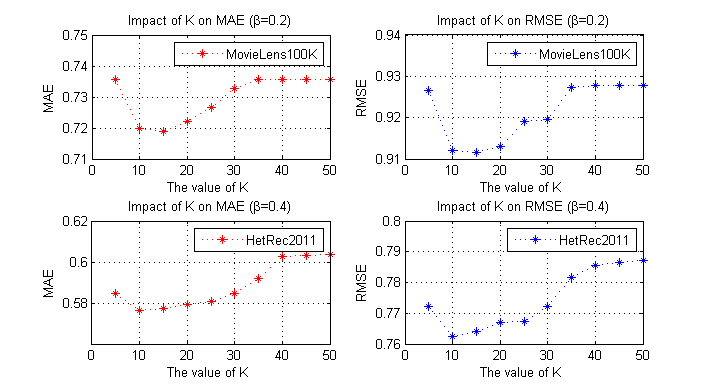}
    \end{center}
    \caption{Impact of Different $K$ on MAE and RMSE.}
    \label{fig:K}
\end{figure}

The dimension of latent feature vectors $K$ is another important parameter in our proposed method. We conduct another group of experiments to assess the impact of parameter $K$ on the recommendation quality of our proposed method by changing $K$ from 5 to 50 with a step of 5. Another parameter $\beta$ is set as $\beta =0.2$ and 0.4 in MovieLens100K and HetRec2011, respectively. The experimental results are plotted in Fig. \ref{fig:K}.


From Fig. \ref{fig:K}, we can clearly see that as $K$ increases, the values of MAE decrease at first, and then begin to increase. Based on the intuition that the greater value of $K$, the more preferences that can be represented by latent feature vectors, and hence better recommendations. However, Fig. \ref{fig:K} shows that continually increasing the value of $K$ does not improve the performance after the dimension of latent feature vector surpasses a certain threshold like 10-dimension on Movielens100K and 20-dimension on HetRec2011. The possible reason is that when $K$ arrives at a specific threshold, the latent user and item feature vectors are enough to characterize the preferences of users or items, and continually increasing $K$ will introduce much noise into the objective function, resulting in degrading recommendation quality.

Our recommendation approach gains the best recommendation quality when $K=15$ and $K=10$ on MovieLens100K and HetRec2011, respectively.

\subsection{Performance on Cold Start Items}
\label{sec:coldItem}
The principle purpose of our proposed approach is to deal with cold start item issue in recommender systems. Although many research work has explored the cold start problem, most of the work focuses on the cold start user problem and ignores the cold start item problem. For example, social networks based recommendation approaches combine social relations between users to solve the cold start user problem. Moveover, as mentioned in Section \ref{sec:dataset}, the cold start item problem is more serious than the cold start user problem in MovieLens100K and HetRec2011. For instance, if we take users who have rated less than 20 items as cold start users, no users are cold start user in MovieLens100K and HetRec2011. In contrast, if we consider items which are rated by users less than 20 times as cold start items, 48.37\% and 44.17\% are cold start items in MovieLens100K and HetRec2011, respectively.

To evaluate the effectiveness of our recommendation approach on coping with the cold start item problem, we firstly group items according to the number of observed ratings on items in the training set, and then compare MAE and RMSE of different item groups with other baseline approaches.

\begin{figure}[!h]
  \centering
  \subfigure[Distribution of Items in Training Dataset of MovieLens100K ]{
    \label{fig:subfig_distribution_100K} 
    \includegraphics[width=3.5in]{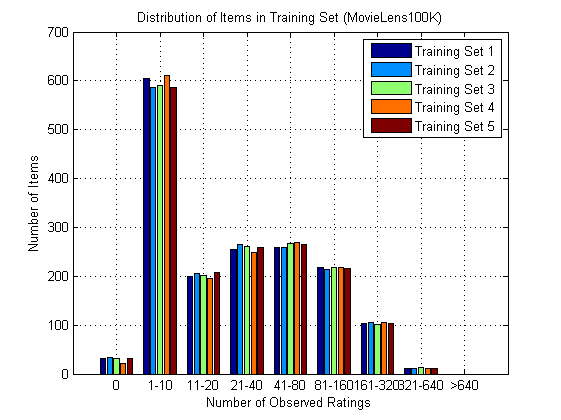}}
  \hspace{0.1in}
  \subfigure[Distribution of Items in Training Dataset of HetRec2011]{
    \label{fig:subfig_distribution_100K} 
    \includegraphics[width=3.5in]{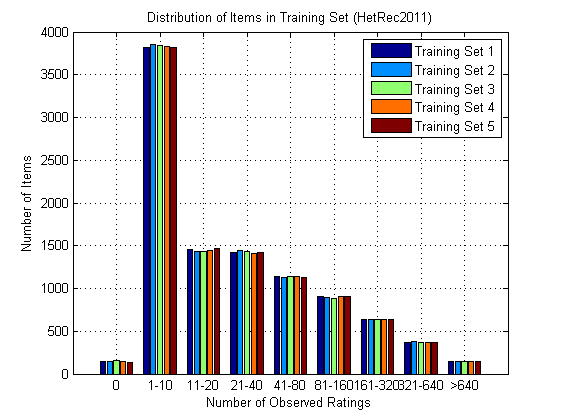}}
  \caption{Distribution of Items in Training Datasets }
  \label{fig:distribution} 
\end{figure}

The distributions of items in each training set for both data sets are depicted in Fig. \ref{fig:distribution}, in which the X-axis shows item groups categories as ``1-10``, ``11-20``, ``21-40``, ``41-80``, ``81-160``, ``161-320`` , ``321-640`` and ``$>$640`` and the Y-axis displays the number of items that are rated the corresponding times. For example, for MovieLens100K data set, there are around 600 items, for which the number of observed ratings in each training set is in the range of [1-10]. Meanwhile, there are around 3800 similar items in the training set of HetRec2011. In this group of experiments, we set $K=10$ and $\beta=0.1$ for both MovieLens100K and HetRec2011 data sets.

\begin{figure}[!h]
  \centering
  \subfigure[ MAE Comparison on Different Item Groups (MovieLens100K)]{
    \label{fig:subfig_coldstart_100K} 
    \includegraphics[width=3.5in]{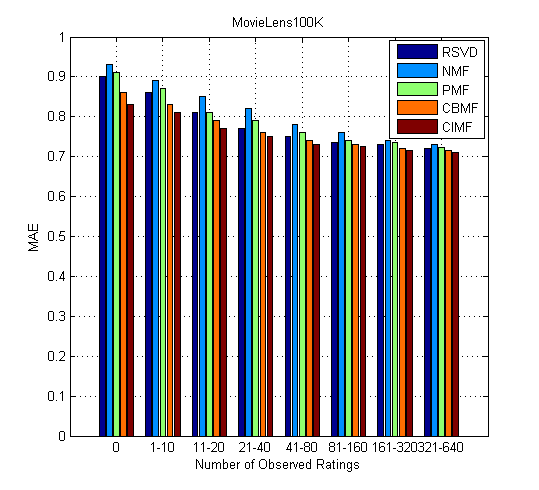}}
  \hspace{0.1in}
  \subfigure[MAE Comparison on Different Item Groups (HetRec2011)]{
    \label{fig:subfig_coldstart_Het} 
    \includegraphics[width=3.5in]{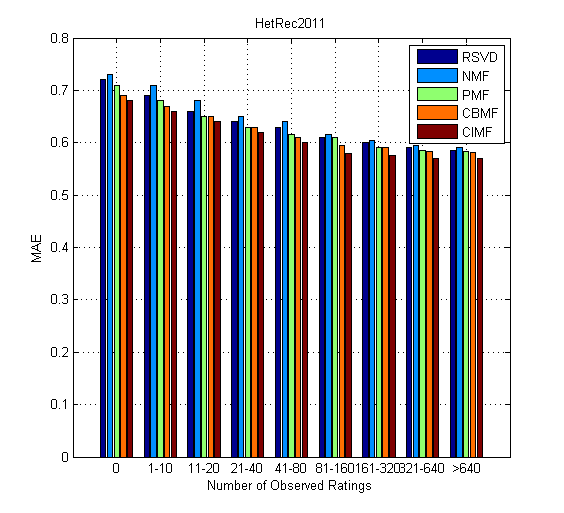}}
  \caption{Performance Comparison on Different Item Groups }
  \label{fig:coldstart} 
\end{figure}

The experimental results are shown in Fig. \ref{fig:coldstart}. Fig. \ref{fig:coldstart} shows that our proposed \emph{IEFM} is able to generate better recommendations than other algorithms, especially for the items with few observed ratings. In terms of MAE, the improvement of our approach for the second category items, i.e., items that are rated from 1 to 10 times, is 5.5\% over RSVD on MovieLens100K and 6.3\% on HetRec2011. As more observed ratings are given, the improvement of our proposed approach gradually reduces, and all compared methods achieve similar performance. These observations indicate that our proposed recommendation algorithm can cope with cold start item problem more effectively than other state-of-art techniques. We argue that the main reason for the improvement is the consideration of the item attribute information as well as the adoption of the coupled object similarity measure to capture the relationships among items in our proposed recommendation algorithm.

\subsection{Impact of Size of Neighborhood of Item}
\label{sec:neighbor}
In this paper, the size of neighborhood for each item is the final control parameter that affects the performance of our proposed approach since two latent item feature vectors are assumed to be close if these two items are similar according to their item attribution contents. In other words, the latent feature vectors of items depend on the feature vectors of their neighbors, especially for those cold start items, the degrees of dependency are greater than those of items that have many observed ratings. To explore the impact of this control parameter on our recommendation algorithm, we vary the numbers of similar neighbors and observe the according changes of recommendation quality. We set K = 10 and $\beta=0.1$ for both MovieLens100K and HetRec2011 data sets.

\begin{figure}[!h]
    \begin{center}
        \includegraphics[scale=.55]{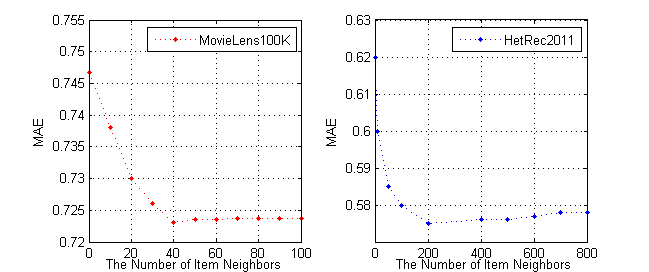}
    \end{center}
    \caption{Impact of Size of Neighborhood on MAE.}
    \label{fig:size_of_neighbor}
\end{figure}

The experimental results are shown in Fig. \ref{fig:size_of_neighbor}. We can observe that the size of neighborhood does have significantly effects on the recommendation quality of our proposed approach for both data sets. Our recommendation approach achieves the best performance when the size of neighborhood is around 40 on MoiveLens100K, while the optimal value of the size of neighborhood on HetRec2011 approximates 200 which is larger than the corresponding value in MovieLens100K. This is primarily because that HetRec2011 contains more items than MovieLens100K. Hence, an item of HetRec2011 generally has more neighbors than items of MovieLens100K. Secondly, HetRec2011 includes more attribution information than MovieLens100K, which can be used to generate more accurate similarity with the coupled object similarity measure. As a result, for HetRec2011, our proposed approach depends more on item neighbors to learn the latent item feature vectors.

We make the following conclusions from the above experimental evaluation.
First, the incorporation of item attribute information is effective in improving the traditional matrix factorization methods, which completely discard additional item content information and only utilize user-item rating matrix to learn the preferences of users and items.
Second, compared with the \emph{SMS }used in \emph{CIMF}, which is relatively rough and fails to capture the relationship between items, the \emph{COS} is more accurate than the \emph{SMS} in capturing the genuine relationship between two categorical items and hence helps recommender systems generate better recommendations for users. Third, our proposed approach can effectively cope with the cold start item problem by keeping the latent item feature vectors of cold start item as close as possible to the latent feature vectors of their neighbors. Finally, when more item attribute information is available, our proposed method can find more reliable neighbors for target items by leveraging the coupled object similarity, resulting in generating better recommendations for users. Hence, lacking of item attribute information would limit the accuracy of our proposed recommendation algorithm.

\section{Conclusion and Future Work}
\label{sec:conclusion}
Recommender systems play an important role in e-commerce for both users and businesses due to the huge
volumes of information on the Web. It provides personalized services for users and promotes more revenues for businesses. In this paper, we propose attributes coupling based item enhanced matrix factorization method  by incorporating item attribute information into matrix factorization technique as well as adapting the coupled object similarity to capture the relationship between items. Item attribute information is formed as an item relationship regularization term to regularize the process of matrix factorization and makes two item-specific latent feature vectors as similar as possible if the two items have similar attribute content. More specially, we adapt the coupled object similarity to capture the genuine relationship between two categorical items, and hence these reliable item neighbors can be leveraged to better characterize the preferences of items. Experimental results on two real data sets show that our proposed method outperforms state-of-the-art recommendation algorithms, such as \emph{RSVD}, \emph{NMF}, \emph{PMF} and \emph{CBMF}.

At present, the available public data sets only contain a small portion of item attribute information and even some popular data sets don't have any related information about items' attributes. For instance, the Netflix contains no item attribute information and MovieLens100K only contains genre information. In the future, we plan to extract more attribute information of item to improve our proposed method. For example, movies' production companies may contribute to the higher values of rating for some users who always tend to those movies that produced by the famous movie production companies, such as, Twentieth Century Fox,  Columbia Pictures Corp. and Warner Bros etc. The more available item attribute information will help  increase the recommendation quality of our propose method.

Moveover, we only constrain item latent feature vectors by using item
attribute information without considering the user social networking relations. In the future, we plan to investigate whether social networking relations are useful for our proposed method to improve the recommendation quality.

Furthermore, although the process of computing similarities among items is offline, the cost of computing similarities measured by the coupled object similarity is expensive, whose time complexity is $O(D^{2}\rho^{3})$. In the future, we plan to investigate how to reduce the time complexity of coupled object similarity measure at the same time keep its advantage and how to use parallel computing method, e.g, MapReduce, to speed up the process of computing the coupled object similarities among items.

\ifCLASSOPTIONcompsoc
  \section*{Acknowledgments}
  The authors would like to thank the  anonymous referees  and the editor
  for their helpful comments and suggestions.
%
%
%

\else
  \section*{Acknowledgment}

\fi

\ifCLASSOPTIONcaptionsoff
  \newpage
\fi



%
\bibliographystyle{IEEEtran}
\bibliography{pakdd}
\end{document}